\documentclass{elsart5}

\newcommand{\paragr}[1]{\textit{#1}}

\usepackage{graphicx}

\usepackage{amssymb}

\begin{document}

\begin{frontmatter}
\title{Ballistic magnetoresistance in small-size carbon nanotubes devices}
\author[aff1]{S. Krompiewski}
\author[aff2]{Gianaurelio Cuniberti}
\address[aff1]{Institute of Molecular Physics, Polish Academy of Sciences,
M.~Smoluchowskiego 17, 60-179 Pozna$\acute{n}$, Poland }
\address[aff2]{Institute for Theoretical Physics, University of Regensburg,
D-93040 Regensburg, Germany}

\begin{abstract}
We theoretically study the magnetoresistance of single wall carbon
nanotubes (SWCNTs) in the ballistic transport regime, using a
standard tight-binding approach. The main attention is directed to
spin-polarized electrical transport in the presence of either
axial or perpendicular magnetic field. The method takes into
account both Zeeman splitting as well as size and chirality
effects. These factors (along with a broadening of energy levels
due to a strong nanotube/electrode coupling) lead, in ultra small
SWCNTs, to serious modifications in profile of the Aharonov-Bohm
oscillations. Other noteworthy findings are that in the parallel
configuration (axial magnetic field) the ballistic
magnetoconductance is negative (positive) for armchair
(semiconducting zigzag) nanotubes, whereas in the perpendicular
configuration the magnetoresistance is nearly zero both for
armchair and zigzag SWCNTs.
\end{abstract}


\end{frontmatter}

\paragr{Introduction}
Electron properties of single wall carbon nanotubes (SWCNTs)
depend on their chirality, \textit{i.e.}\ on how a graphene strip
is rolled into a cylinder. The chirality is uniquely determined by
the so-called wrapping vector (\emph{n,m}) defined in terms of
graphene primitive vectors. In this study we restrict ourselves to
(armchair) (\emph{n,n})-type nanotubes and (zigzag)
(\emph{n,0})-type ones. It is well-known that the former are
metallic, whereas the latter can be semiconducting unless \emph{n}
is a multiple of 3. Yet it was predicted theoretically (see
Ref.~\cite{Ando} and references therein) and confirmed
experimentally~\cite{Coskun,Zaric} that a nominally semiconducting
SWCNT could be turned into a metallic one (and vice versa) upon
application of an axial magnetic field. Here we study the effect
of reduction of the nanotube size on the conductance, and show
that nanotubes whose length and diameter are close to each other
cannot serve as efficient magneto-electrical switches.
Additionally, we calculate the magnetic field induced spin current
(in the linear response regime) and show that it depends
critically both on the chirality of SWCNTs and mutual orientation
of the external field and the nanotube axis.
\begin{figure}[b]
\centerline{\includegraphics[scale
=0.7]{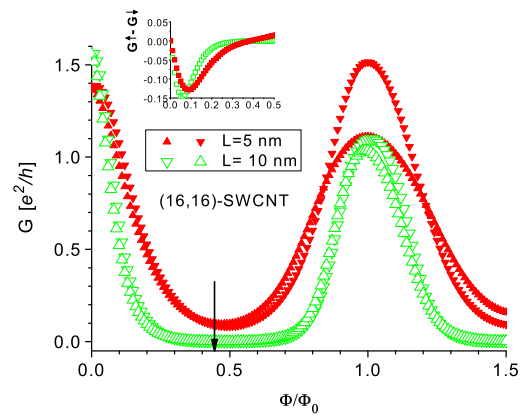}} \caption{\label{fig-1} Axial
magnetic field magnetoresistance of (16,16)-SWCNTs \emph{circa}
5~and 10~nm in length ($L$), against the normalized magnetic flux.
The inset shows the difference in conductance of majority- and
minority-spin carriers.}
\end{figure}

\paragr{Method and Results}
Our starting point is a model, extensively described in
Ref.~\cite{pss05,KGC}, which proved to successfully describe
electrical transport through SWCNTs end-contacted to metal
electrodes.
\begin{figure}[t]
\centerline{\includegraphics[scale
=0.7]{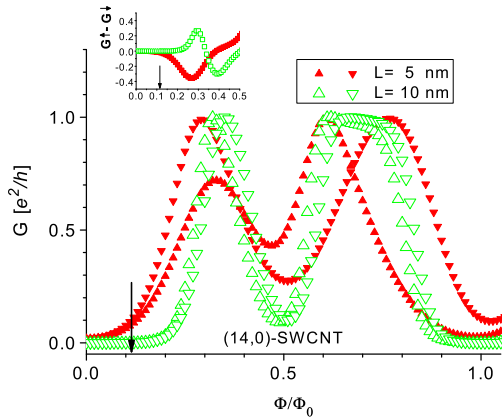}} \caption{\label{fig-2}As
Fig.~1 but for a zigzag (14,0) SWCNT. The inset shows that at
magnetic fields within reach of a lab, there is practically no
conductance spin-splitting.}
\end{figure}
In this work, we additionally introduce an external magnetic field
($B$) as (i) a Zeeman splitting ($\pm \frac{1}{2} g \mu_B B$) to
the diagonal matrix elements of the tight-binding hamiltonian, and
(ii) the standard Peierls substitution. The latter is implemented
by renormalizing the off-diagonal hopping parameters as $t(B)
\longrightarrow t \;\exp( \textrm{i}\ 2 \pi \varphi /\Phi_0 )$,
where $\Phi_0=h/e$ denotes the flux quantum. The hopping integral
is $t =2.7$~eV, and the general expression for the flux $\varphi$
reads

\begin{equation} \label{zeta}
\varphi=\cases{ \frac{\Delta x}{C}\Phi, &
$\!\!\!\!\!\!\!\!\!\!\!\!\!\!\!\!\!\!\!$ for parallel fields, \cr
\left(\frac{C}{2 \pi}\right)^2 \frac{B \Delta y}{\Delta x}
\left[\cos \frac{2 \pi
x}{C}-\cos \frac{2 \pi (x+ \Delta x)}{C}\right] , 
\cr &
$\!\!\!\!\!\!\!\!\!\!\!\!\!\!\!\!\!\!\!\!\!\!\!\!\!\!\!\!\!\!\!\!\!\!\!\!
$ for perpendicular fields. \cr}
\end{equation}

Here $\Phi=B \pi (\frac{C}{2 \pi})^2$ is the parallel magnetic
flux piercing a SWCNT's perimeter $C=a \sqrt{n^2+m^2+mn}$, $a=2.49
\AA$ is the graphene lattice constant and
$x$ ($y$)
is a circumferential (axial) coordinate.

Figures~\ref{fig-1}-\ref{fig-2} show the ballistic
magnetoconductance spectra in the parallel configuration for
armchair and zigzag SWCNTs, whereas the respective insets present
spin polarization of the conductance as a function of normalized
magnetic flux $\Phi/\Phi_0$. In all the figures the up- and
down-triangle markers relate to $\uparrow$ and $\downarrow$ spin
contributions to the conductance. Open and closed markers refer to
longer (10 nm) and shorter (5 nm) nanotubes, respectively. For
clarity of presentation, a vertical arrow pointing to the field
value of $B=500 T$ is also shown.

\begin{figure}[t]
\centerline{\includegraphics[scale
=0.73]{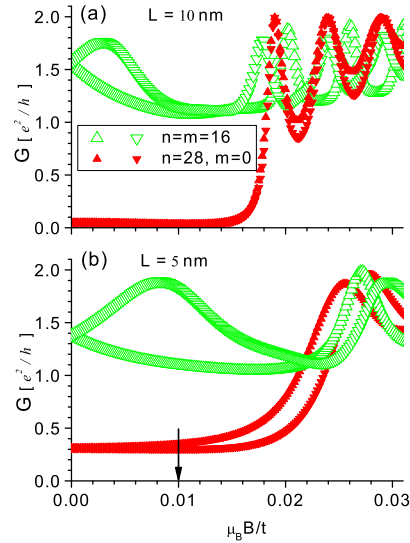}} \caption{\label{fig-3}The
effect of the SWCNT chirality and length ($L$) on the
spin-resolved conductance at perpendicular magnetic field $B$. (a)
$L\sim$~10~nm, (b) $L\sim$~5~nm. Note that the conductance is
strongly spin-split already at relatively small $B$ in the
armchair case in contrast to the zigzag one. Moreover the ultra
short zigzag SWNT in (b) turns out to be conductive even at $B=0$,
due to the fact that its length and diameter are comparable
(resulting in extra broadening of energy levels and closure of the
energy gap).}
\end{figure}

As evident in Fig.~\ref{fig-1}, with increasing lengths the
conductance of armchair tubes decreases to zero at $\Phi = \Phi_0
/ 2$ and $3 \Phi_0 / 2$ approaching the ``$h/e$'' Aharonov-Bohm
periodicity. Remarkably, the magnetoconductance is predominantly
negative and the spin-polarization ($G_{\uparrow}-G_{\downarrow}$)
is quite noticeable although very hardly $L$-dependent. This is in
contrast to what is depicted in Fig.~\ref{fig-2}, for the case of
a semiconducting zigzag tube, which shows no spin-splitting of the
conductance (see Inset) and positive magnetoconductance. Finally
Fig.~\ref{fig-3} shows that, in the perpendicular field
configuration, there is a considerable conductance spin-splitting
for armchair SWCNT. The magnetoresistance is however very weak
($G_{\uparrow}+G_{\downarrow}$ is roughly independent of $B$ at
accessible magnetic fields).

\paragr{Conclusions}
For ultra short SWCNTs (lengths and diameters of comparable size),
the ballistic magnetoconductance at axial magnetic fields is
drastically changed. In particular, instead of distinct gaps in
the conductance spectrum only pseudo-gaps, \textit{i.e.}\ regions
with suppressed but finite conductance occur.  At perpendicular
$B$, in turn, the finite-length effect is still quite pronounced
for zigzag nanotubes, in contrast to armchair ones for which there
are no major changes in the total $G$ \emph{vs}\ $B$ spectra. It
has been also found that both magnetoresistance and electrical
spin-polarized conductance are basically chirality-, diameter- and
length-dependent. However ballistic magnetoresistance practically
vanishes for the perpendicular configuration because its
spin-dependent components compensate each other. As regards the
parallel configuration, the ballistic magnetoresistance is
positive for armchair and negative for zigzag SWCNTs.

\paragr{Acknowledgements}
This work was funded by the EU grant CARDEQ under contract
IST-021285-2, the DFG collaborative research center ``Spin phenomena
in reduced dimensions''(SFB 689/A3) and by the Volkswagen Foundation
under grant No.~I/78~340. SK acknowledges the Pozna{\'n}
Supercomputing and Networking Center for the computing time.

\end{document}